\documentclass[%
 reprint,
 amsmath,amssymb,
 aps,
]{revtex4-2}

\usepackage{graphicx}
\usepackage{dcolumn}
\usepackage{bm}

\usepackage{url}
\usepackage{hyperref}
\begin{document}

\preprint{APS/123-QED}

\title{Quantum Transfer Learning with Adversarial Robustness for Classification of High-Resolution Image Datasets}

\author{Amena Khatun\textsuperscript{1}}
\email{amena.khatun@data61.csiro.au}
\author{Muhammad Usman\textsuperscript{2,3}}

\affiliation{%
\textsuperscript{1}Data61, CSIRO, Dutton Park, QLD 4102, Australia\\
\textsuperscript{2}Data61, CSIRO, Research Way, Clayton, 3168, Victoria, Australia\\
\textsuperscript{3}School of Physics, The University of Melbourne, Parkville, 3010, Victoria, Australia
}%


\begin{abstract}
The application of quantum machine learning to large-scale high-resolution image datasets is not yet possible due to the limited number of qubits and relatively high level of noise in the current generation of quantum devices. In this work, we address this challenge by proposing a quantum transfer learning (QTL) architecture that integrates quantum variational circuits with a classical machine learning network pre-trained on ImageNet dataset. Through a systematic set of simulations over a variety of image datasets such as Ants \& Bees, CIFAR-10, and Road Sign Detection, we demonstrate the superior performance of our QTL approach over classical and quantum machine learning without involving transfer learning. Furthermore, we evaluate the adversarial robustness of QTL architecture with and without adversarial training, confirming that our QTL method is adversarially robust against data manipulation attacks and outperforms classical methods.
\end{abstract}

\maketitle

\begin{figure*}
\begin{center}
\includegraphics[width=0.8\linewidth]{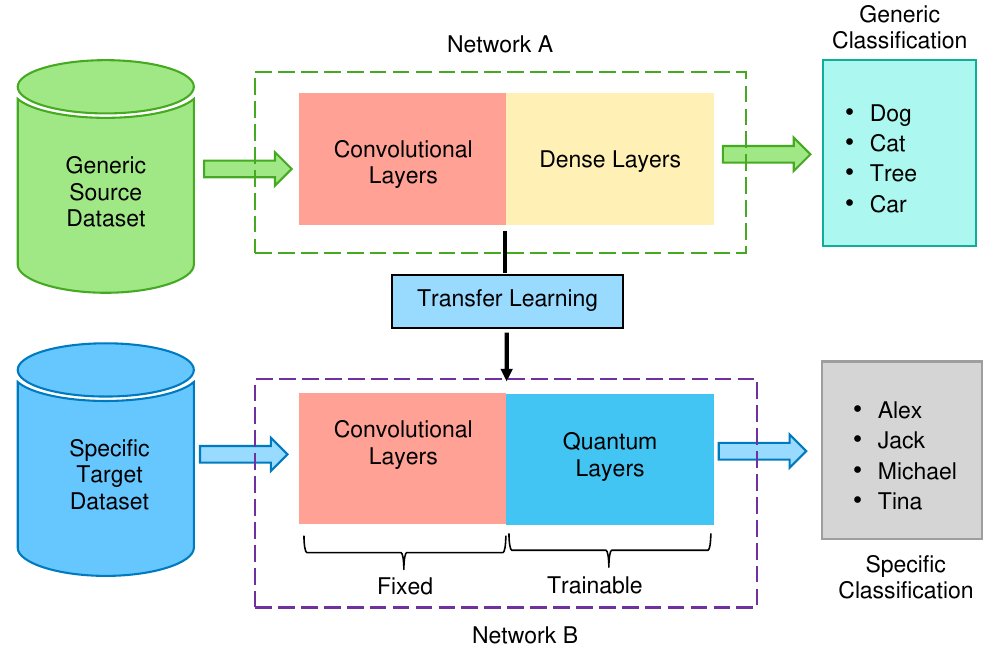}
\end{center}
\caption{Overview of Quantum Transfer Learning (QTL). Network A is trained on a source dataset which is typically large and diverse. The purpose of training Network A is to learn general features, patterns, and representations that are applicable to the broader domain. Here, Network A will learn to recognise basic shapes, textures, and patterns commonly found in images of various animals and objects. Network B is the target model that is used for a different but related task. Instead of starting the training of Network B from scratch, the weights and parameters are initialised and learned by Network A. QTL model performs well on a smaller or completely different dataset compared to the source dataset as it learns features and representations from a source task to initialize a model for the target task. Even if the source and target domains are different, many lower-level features (edges, colors, and textures) and patterns are common to both domains. After initialization, Network B is further trained on the target dataset. During this fine-tuning process, the model adapts its representations to the specific characteristics and requirements of the target task. This fine-tuning allows the model to specialize while retaining the general knowledge from Network A.}
\label{fig:QTL}
\end{figure*}
\section{\label{sec:level1}Introduction}

Machine learning (ML) has gained immense popularity in recent years for autonomously learning from data, leading to its applications in various domains such as self-driving, security and surveillance, face recognition, object classification, behaviour analysis, and anomaly detection. However, as the demand for computational power and efficiency continues to grow, researchers are investigating Quantum Machine Learning (QML) models to address the growing computational demands of ML models \cite{cerezo2022challenges, west2023reflection, ren2022experimental, west2023provably, huang2022quantum, west2023boosted, tsang2023hybrid}. QML employs the principles of quantum mechanics to design fundamentally new machine learning algorithms with the potential to offer computational efficiency and accuracy across a wide range of complex problem-solving tasks. To exploit the benefit of quantum computing and machine learning in a single framework, researchers developed hybrid quantum convolutional neural networks (QCNNs) \cite{cong2019quantum, beer2020training} which have already demonstrated superior performance in object classification \cite{chen2023quantum, hur2022quantum, wang2022development}, particularly in low-parameter scenarios while the real-world datasets are characterized by hundreds of high-dimensional features. To address this challenge and enable the application of QML to high dimensional feature space in complex data sets, the QTL approach has been explored that integrates classical ML with quantum neural network (QNN) in a unified framework \cite{mari2020transfer, otgonbaatar2022quantum, mogalapalli2022classical, azevedo2022quantum, alex2023hybrid}.

Transfer Learning (TL) \cite{bengio2012deep} aims to enhance the training of a new ML model by leveraging a pre-trained reference model that has been previously trained on a distinct but related task using a different dataset. TL is particularly advantageous when training deep learning models that require significant time due to the abundance of data, especially when the features learned in the initial layers have broad applicability across various datasets. In such cases, it is more practical to start with a pre-trained network and fine-tune only a subset of the model's parameters for a specific task, rather than training the entire network from scratch. As an example, suppose a reference neural network is trained on dataset X to solve task A. Now, when confronted with dataset Y for solving task B, instead of starting the training process from scratch, the parameter values such as weights and biases obtained from the earlier layers of the reference neural network initialised as fixed parameters for the new neural network that is dedicated to solving task B using dataset Y. Figure \ref{fig:QTL} illustrates the overview of TL approach. TL models perform well even if the target dataset is completely different than the source dataset as many low-level features such as patterns, edges, colors, textures, etc. which are more general and abstract features might be common for both domains \cite{neyshabur2020being, kornblith2019better}. By employing transfer learning effectively, significant performance improvement is achieved with faster training time.

Despite significant achievement in the performance of QML models for various tasks, handling high-dimensional data in QML models suffers from several challenges: 1) the number of qubits required to represent the entire space of high-dimensional data grows with the dimensionality, however, the available quantum computers have a limited number of qubits, making it challenging to perform computations on high-dimensional data. 2) when the dimensionality of the data increases, the depth of the quantum circuit also increases to process the data leading to the accumulation of noise or errors. 3) Encoding classical data into a quantum state for processing is not straightforward, especially for high-dimensional data. Efficient quantum data encoding is still an open area of research \cite{nakaji2022approximate, larose2020robust, wu2023radio, west2023drastic}, and the selection of encoding impacts the performance of quantum algorithms. To address these challenges, a few QTL approaches \cite{mari2020transfer, otgonbaatar2022quantum, mogalapalli2022classical, azevedo2022quantum, alex2023hybrid} have been developed in the last two years. Most of these approaches focused on developing QTL methods for image classification with various applications such as trash detection \cite{mogalapalli2022trash}, spotting cracks from image data \cite{alex2023hybrid}, detecting tuberculosis \cite{mogalapalli2022classical}, and breast cancer detection \cite{azevedo2022quantum}. Qi \textit{et. al.} \cite{qi2022classical} introduced a QTL approach to recognise the spoken command. While these QTL approaches show promising results, these studies lack a comprehensive comparison between QTL and classical TL methods. Without such comparisons, it is difficult to understand if the observed performance gains in QTL are outperforming the classical approaches. Another notable limitation of the current QTL approaches is the lack of research on their vulnerability to adversarial attacks. While adversarial attacks and their defense mechanisms have been extensively studied in classical machine learning \cite{szegedy2013intriguing, huang2011adversarial, DBLP:journals/corr/GoodfellowSS14, DBLP:conf/iclr/KurakinGB17a, ilyas2018black, DBLP:conf/iclr/TjengXT19}, their impact on quantum models, especially in the context of TL, remains largely unexplored. Adversarial attacks are a serious threat to the security and robustness of ML models, as they can lead to incorrect or malicious predictions. It is also important to understand how adversarial perturbations affect the performance of QTL models, otherwise, the reliability of quantum models could be compromised. Thus, investigating the resilience of QTL methods against adversarial attacks is essential to ensure the safety and effectiveness of these models in real-world scenarios.

To overcome the above-mentioned limitations of the current literature and to address knowledge gaps, in this paper, we propose a QTL approach that not only demonstrates the benefits of quantum computing in the context of TL but also establishes a direct comparison between quantum and classical TL methodologies. By benchmarking our QTL approach against classical TL and non-transfer learning scenarios, we provide clear insight into the advantages of QTL and its performance improvements. By performing a systematic set of classical and quantum simulations, we quantitatively demonstrate the improved accuracy of our QTL model over the classical TL technique. We also investigate the susceptibility of the proposed QTL model to adversarial attacks to determine the security and reliability of the model. In response to the vulnerabilities identified through adversarial testing, our results show that: the proposed QTL model exhibits enhanced robustness against adversarial attacks, forming the basis for quantum adversarial transfer learning (QATL). By iteratively exposing our model to adversarial examples during training, we enforce the model to learn and adapt to adversarial perturbations, thereby enhancing its resistance to malicious inputs and ensuring more reliable predictions in real-world scenarios.

\section{\label{sec:level2}Literature Review}

In this section, we briefly summarise the related research in quantum-classical hybrid convolutional neural networks, quantum-classical transfer learning, quantum adversarial attack and robustness of QML, and quantum adversarial training.

\subsubsection{Quantum-Classical Hybrid Convolutional Neural Networks (QCHCNNs)}
The success of machine learning algorithms depends on the computational power, which grows proportionally with the increasing volume of data. Quantum computers promise to offer immense computational power and therefore it is natural to integrate machine learning with quantum computing to develop new QML models. Thus, researchers introduced QML algorithms \cite{rebentrost2014quantum, biamonte2017quantum, lloyd2014quantum, lloyd2013quantum}, demonstrating their potential to achieve remarkable speedups compared to their classical counterparts. Motivated by the potential power of QML, several QCNNs methods have been developed in recent years for various tasks \cite{cong2019quantum, beer2020training, chen2023quantum, kerenidis2019q, henderson2020quanvolutional, liu2021hybrid, hur2022quantum, bengio2012deep, wang2022development}. In Ref. \cite{cong2019quantum}, a QCNN is proposed, which efficiently trains and operates on near-term quantum devices by utilising O(log(N)) variational parameters for N qubits. The proposed method incorporates both the multi-scale entanglement and quantum error correction techniques. Ref. \cite{kerenidis2019q} introduces a quantum algorithm for clustering, analogous to classical k-means which provides an exponential speedup compared to classical k-means for well-clusterable datasets, making it a promising tool for quantum machine learning applications. In Ref. \cite{maccormack2022branching}, a branching quantum convolutional neural network (bQCNN) is proposed to make use of the global information obtained from all the qubits that are measured at the pooling layer instead of only using the local information. This approach allows the network to consider the collective outcomes of all measured qubits, enabling it to capture long-range correlations between qubits. By doing so, bQCNN potentially offers improved performance in detecting patterns and features that span larger distances in the quantum data. Ref. \cite{chen2022quantum} introduces a novel hybrid QCNN framework to demonstrate the quantum advantage compared to classical algorithms for the classification of High-Energy Physics (HEP) events. Despite having a similar number of parameters in both QCNN and classical CNNs, the hybrid QCNN demonstrates a faster learning rate and achieves better testing accuracy with fewer training epochs.

Other research focused on developing QCNN approach for image classification as image classification is one of the most useful machine learning tasks with diverse applications in areas such as autonomous driving \cite{chen2017multi, maturana2015voxnet}, defense (target identification, surveillance, and threat assessment), healthcare \cite{mckinney2020international, esteva2017dermatologist}, and surveillance and security \cite{schroff2015facenet, taigman2014deepface}. Due to the wide range of real-world applications of image classification, several QCNNs have been developed for image classification tasks in recent years, such as, Quanvolutional neural network (QNN) is introduced in \cite{henderson2020quanvolutional}, where a transformational layer is introduced to locally transform input data using random quantum circuits, extracting meaningful features for classification purposes. The proposed QNN is evaluated on the MNIST dataset and shows that QNN models outperform classical CNNs in terms of test accuracy and training speed. Ref. \cite{chen2023quantum} introduces Multi-scale Entanglement Renormalization Ansatz (MERA) and Box-counting based fractal feature extraction methods within the QCNN framework for binary image classification. The proposed new feature extraction methods outperform the classical CNN in terms of recognition accuracy for the breast cancer dataset. A fully parameterized QCNN is developed in \cite{hur2022quantum} for classical data classification considering two-qubit interactions and shallow-depth quantum circuits, making it suitable for NISQ devices. Extensive benchmarking on MNIST and Fashion MNIST datasets shows that QCNN achieves high classification accuracy despite having a small number of parameters.

\subsubsection{Quantum Transfer Learning}
Despite notable progress towards the development of hybrid QCNNs for image recognition, their application for complex datasets is still a challenge due to the limited number of qubits and relatively high level of noise. To overcome these challenges, the recent work has focused on TL and developed QTL approaches \cite{mari2020transfer, otgonbaatar2022quantum, mogalapalli2022classical, azevedo2022quantum, alex2023hybrid} for various ML tasks. In Ref. \cite{mari2020transfer}, the concept of TL is adopted in the context of hybrid QNN that combines classical and quantum elements. The authors propose a variety of implementations of hybrid transfer learning, with a focus on modifying and augmenting a pre-trained classical network using a final quantum variational circuit (QVC). This approach is advantageous in the era of intermediate-scale quantum technology as it enables efficient pre-processing of high-dimensional images with classical neural networks while using the power of a quantum processor to extract essential and informative features. Ref. \cite{otgonbaatar2022quantum} employed QTL, combining multi-qubit QML networks and deep convolutional neural networks to extract features from small, high-dimensional datasets. This method uses real-valued amplitude and strong entanglement in QML networks, and the classification performance is evaluated on Eurostat and synthetic datasets. The QML model is trained via QTL in real-world, small, and high-dimensional images (256 × 256 × 3). Ref. \cite{mogalapalli2022classical} proposes a method where pre-trained classical feature extractors such as DenseNet169 \cite{huang2017densely}, VGG19 \cite{simonyan2014very}, or AlexNet \cite{krizhevsky2012imagenet} are combined with a quantum circuit as a classifier for three distinct image classification tasks: detecting organic and recyclable materials from Trash dataset, identifying tuberculosis from chest X-ray images, and spotting cracks in concrete crack images. The results demonstrate that different pre-trained networks (DenseNet, AlexNet, VGG19) perform better on specific datasets, with DenseNet achieving high accuracy on trash images, AlexNet on TB detection, and VGG19 on crack detection. Due to the limited number of qubits, the quantum circuit was used only for the final classification. QTL is also proposed in Ref. \cite{azevedo2022quantum} for breast cancer detection to classify full-image mammograms using pre-trained CNN with the aid of quantum-enhanced TL. Ref. \cite{alex2023hybrid} followed a similar concept and architecture as proposed in Ref. \cite{mari2020transfer}, however, they aim to detect the cracks in grayscale images. Ref. \cite{kim2023classical} propose another Classical-to-quantum TL approach and perform numerical simulations with different sets of quantum convolution and pooling operations for MNIST data classification. Initially, a classical CNN is trained on the Fashion-MNIST dataset. Although the source dataset (FMNIST) is very different than the target dataset (MNIST), the shared lower layers learned by the classical CNN capture low-level features such as edges, shapes, textures, colors, which are more general and abstract features. Hence, when applying transfer learning to a target dataset that is different from the source dataset, these lower layers serve a crucial role and reduce the need for training from scratch on MNIST and potentially improve the performance of QCNN models on the target dataset. While all the above-mentioned QTL research focused on developing QTL methods for image classification, Ref. \cite{qi2022classical} introduces a hybrid classical-to-quantum
TL method for spoken command recognition (SCR). The proposed architecture consisted of two parts, a one-dimensional classical CNN for speech feature extraction and a quantum part that is built on QVC with fewer parameters.

Based on the literature survey, we find that the existing work has made significant progress in addressing the challenges posed by limited qubits and high levels of noise in quantum circuit implementations for image classification. Several QTL approaches have been proposed, combining classical and quantum elements to leverage TL in the quantum domain. However, it is worth noting that none of the above-mentioned studies investigate the effect of the model without transfer learning, i.e., when there is no pre-trained network involved. Moreover, the comparison of QTL results with their classical counterparts is limited in these studies. Although existing QTL methods show promising potential in enhancing image classification tasks, further exploration and comprehensive comparisons between QTL and classical methods are necessary for a better understanding of the benefits and limitations of QTL in real-world applications. Moreover, most of the existing QTL approaches are limited to lower-resolution image datasets such as Eurosat (64$\times$64$\times$3) \cite{otgonbaatar2022quantum}, UC Merced Land Use (256$\times$256$\times$3) \cite{otgonbaatar2022quantum}, MNIST (28$\times$28$\times$3) \cite{kim2023classical} for image classification, while our proposed method considered high-resolution image datasets, Ants \& Bees (768$\times$512$\times$3), and Road Sign Detection (1024$\times$1024$\times$3) for image classification.

\subsubsection{Quantum Adversarial Attack and Robustness of QML}
It is well known that even the state-of-the-art classical ML models demonstrate vulnerability to small, carefully crafted perturbations to just a few pixels. These adversarial perturbations cause misclassification by the well-performing classifiers. This raises questions about the vulnerabilities of high-performing ML classifiers \cite{szegedy2013intriguing, huang2011adversarial, DBLP:journals/corr/GoodfellowSS14, DBLP:conf/iclr/KurakinGB17a, ilyas2018black, DBLP:conf/iclr/TjengXT19} and the security risks to ML applications. Recent work has demonstrated that quantum learning systems, similar to classical classifiers based on classical neural networks, are vulnerable to adversarial examples \cite{szegedy2013intriguing, 43405}, irrespective of whether the input data is classical or quantum in nature \cite{liao2021robust, lu2020quantum, anand2021noise, guan2021robustness, gong2022enhancing, geng2023hybrid}. Even quantum classifiers achieving nearly state-of-the-art accuracy can be easily fooled by the adversarial examples generated by introducing imperceptible perturbations to the original legitimate samples. 

In Ref. \cite{lu2020quantum}, quantum machine learning and its susceptibility to adversarial scenarios are explored. This research reveals that even subtle modifications to the original input data cause high-accuracy quantum classifiers to falter under manipulation. This investigation encompasses a wide range of scenarios, including the classification of handwritten digits in the MNIST dataset, identification of the phases of matter, and classification of quantum data. This study considered both the white-box and black-box attack scenarios. In the
white-box scenario, where the attacker has full information about the
learned model and the learning algorithm, the adversarial perturbations are generated for quantum classifiers. Even only a small amount of noise causes the accuracy of quantum classifiers to decrease significantly. 
To further verify the vulnerability of the quantum classifier, this work performs simulations under a black-box setting, assuming that the attacker has limited or no access to the internal architecture or the classifiers. For the black-box setting, adversarial examples are generated for classical classifiers and transferred these attacks to quantum classifiers. The results demonstrate that quantum classifiers are vulnerable to the adversarial examples generated for classical classifies due to the transferability properties of adversarial examples.

In another recent study Ref. \cite{west2023towards}, the performance of QVCs, CNNs, and ResNet-18 was systematically compared in both white-box and black-box scenarios. The findings revealed a notable resilience of quantum classifiers when subjected to attacks generated for classical classifiers. This discovery stands in contrast to the results reported in a prior study in Ref. \cite{lu2020quantum}, highlighting a distinct observation. The authors in Ref. \cite{west2023towards} emphasis that quantum classifiers exhibit greater resilience because the QVCs learn a distinct yet highly meaningful set of features compared to classical classifiers.
.
Although several QML approaches have been developed in recent years to ensure the security and reliability of QML applications, translating these theoretical concepts into practical implementations that can effectively defend QML models against adversarial attacks remains an open challenge.

\subsubsection{Quantum Adversarial Training}
The importance of enhancing the robustness of quantum classifiers is crucial to ensure the security and reliability of QML applications. Only a limited amount of work \cite{lu2020quantum, ren2022experimental, west2023benchmarking} has focused on employing adversarial training (AT), inspired by its success in classical ML. AT has been shown as one of the key defense strategies for classical ML models against adversarial attacks.

Ref. \cite{lu2020quantum} demonstrates that quantum classifiers are as vulnerable to adversarial attacks as classical classifiers. This study brought the idea of AT in QML and significant performance improvement of the quantum classifier is achieved against classical adversarial attack. However, the AT is only evaluated on the downscaled MNIST dataset for simple binary classification and only BIM (basic iterative method) attack \cite{feinman2017detecting} is considered to generate adversarial examples. To clarify the robustness of the model, more complex and high-dimensional datasets need to be evaluated on multi-class classification. While Ref. \cite{lu2020quantum} emphasises that AT helps to improve the robustness of quantum classifiers against classical attacks, another study Ref. \cite{west2023benchmarking} presents a comprehensive framework for evaluating the robustness of quantum and classical ML networks and compares the performance of QVC with classical CNNs and ResNet-18 in the presence of both classical and quantum adversarial attacks. This work demonstrates that adversarial training provides significant benefits for classical ML networks against classical attacks. However, when adversarial training is applied to QML networks, particularly on QVC, the robustness of the quantum classifier against classical attacks improves to some extent. This improvement, however, is negligible compared to classical ML performance. This study emphasises that AT does not enhance the robustness of QML against classical attacks, as QML is already resilient to such attacks. There are significant differences between the methods proposed in Ref. \cite{west2023benchmarking} and Ref. \cite{lu2020quantum} as Ref. \cite{west2023benchmarking} performs multi-class classification with different types of adversarial attacks such as PGD \cite{madry2017towards}, FGSM \cite{43405}, and AutoAttack \cite{croce2020reliable} while Ref. \cite{lu2020quantum} perform binary classification under BIM attack only. Overall, the impact of adversarial training on the robustness of QML is still an open research question which will require future work to perform a systematic investigation of a variety of QML architectures as well as datasets.


\section{Methodology}
\subsection{Knowledge Transfer from Classical CNNs to Quantum Classical Hybrid Convolutional
Neural Network (QCHCNNs)}
\begin{figure*}
\begin{center}
\includegraphics[width=1.0\linewidth]{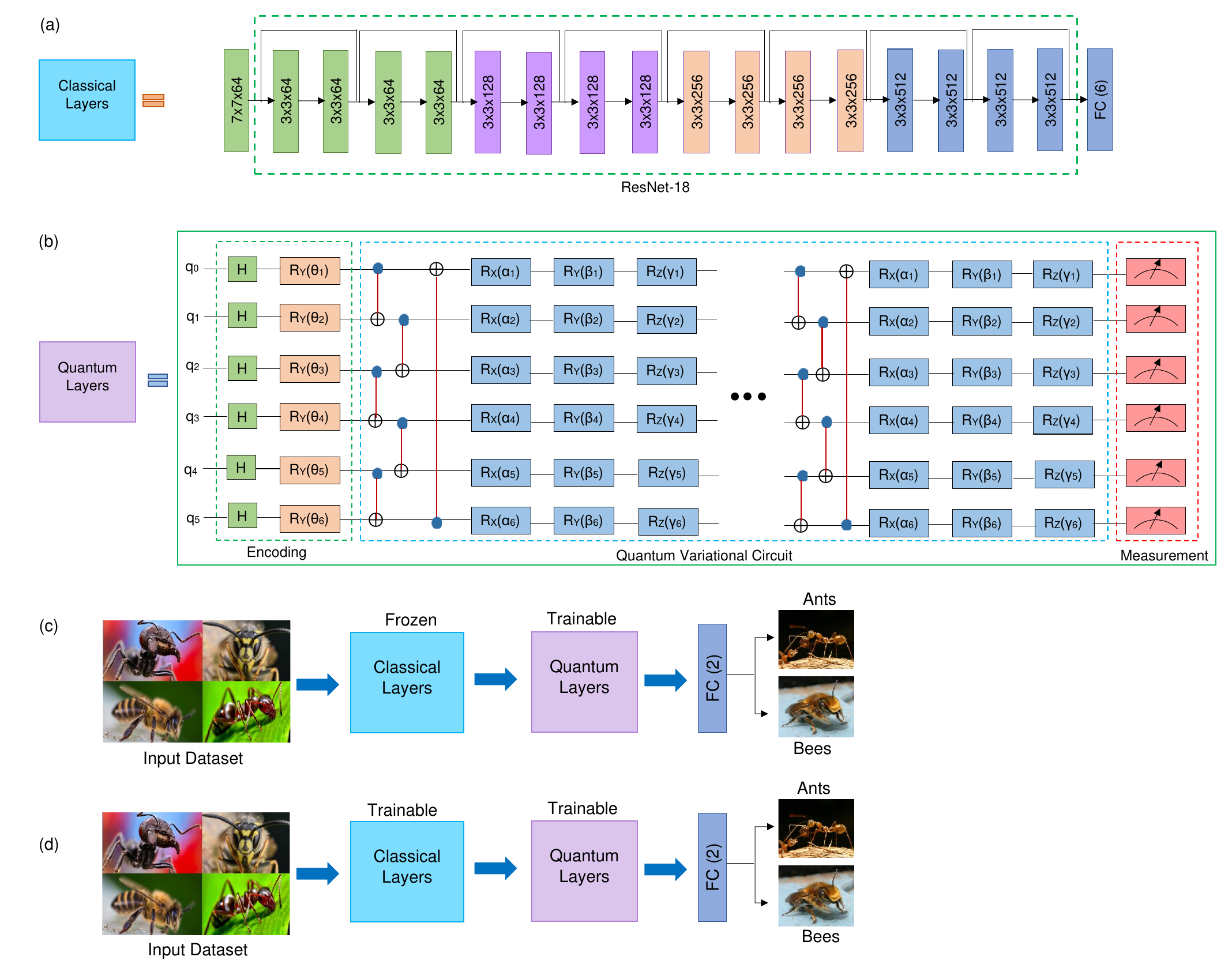}
\end{center}
   \caption{The architecture of the proposed QTL method. The pre-trained network is ResNet-18 which consists of 18 layers. In the proposed QTL architecture (a) represents the overview of the proposed QTL framework, where the input data is fed to the classical layers for pre-processing. All the layers of ResNet-18 are frozen except the last dense layer during the entire training phase. ResNet-18 is considered only as a feature extractor where the extracted features are flattened and fed into a fully connected layer. The outout of the classical network is then fed into the trainable quantum layers. Finally, the measured output is fed into a fully connected layer for final prediction. (b) represents trainable quantum layers and classical layers. Trainable quantum layers consist of data encoding, QVC, and a measurement layer. Here, QVC consists of a combination of CNOT gates and single-qubit adjustable rotation gates $(R_X, R_Y, R_Z)$. The classical layer is ResNet-18 which consists of 18 layers. Each layer consists of two blocks and each block consists of two weight layers with a skip connection that is connected to the output of the next weight layer. (c) represents the overview of the proposed QTL framework, where the input data is fed to the classical layers for pre-processing. All the layers of ResNet-18 are frozen except the last dense layer during the entire training phase. ResNet-18 is considered only as a feature extractor where the extracted features are flattened and fed into a fully connected layer. The outout of the classical network is then fed into the trainable quantum layers. Finally, the measured output is fed into a fully connected layer for final prediction. (d) represents the architecture of quantum-classical hybrid convolutional neural network without transfer learning where all the convolutional layers are also trainable in this case.}
\label{fig:QTL_Architecture}    
\end{figure*}
The proposed QTL framework adopts the concepts of TL to train Quantum-Classical Hybrid Convolutional Neural Networks (QCHCNNs) to learn the features of the large training set and evaluate the results on a specific target dataset such as Ants \& Bees, Road sign detection, and CIFAR-10 datasets. In the proposed architecture, ResNet, also known as a deep residual network \cite{he2016deep} is used as a pre-trained network from which the knowledge is transferred from the source domain to the target domain. Due to its residual connection, ResNet helps to transfer the information within the network in a better way and also propagates the gradients directly through the network. More specifically, the residual connection of ResNet takes the global feature into account by skipping the connections that lead to less training time and optimal tuning of the network layers. However, the degradation issue arises when the network is gets deeper, i.e., when the depth is increasing, the accuracy gets saturated followed by a rapid degradation. This problem is addressed by the deep ResNet \cite{he2016deep} where the deeper layers learn a residual mapping rather than relying on the assumption that a few consecutively stacked layers will precisely capture the intended underlying mapping. 

In the proposed architecture, ResNet-18 is used as the pre-trained network which was trained on a larger dataset, Imagenet \cite{russakovsky2015imagenet} that won 1st place in the ILSVRC (ImageNet Large Scale Visual Recognition Challenge)
2015 classification competition. ImageNet is successfully used in transfer learning and has been proven to be competent in pre-training the model for a specific task in classical ML \cite{neyshabur2020being, kornblith2019better, huh2016makes}. This dataset contains 1,281,167 training images, 50,000 validation images, and 100,000 test images of 1000 classes. The architecture of ResNet-18 is illustrated in Figure \ref{fig:QTL_Architecture}(a). At first, the input images go through a convolutional layer with a 7$\times$7 filter and a stride of 2, followed by max-pooling with a 3$\times$3 filter and a stride of 2. This initial processing helps reduce the dimensions of the input and capture initial features.  Each layer consists of two blocks and each block consists of two weight layers with a skip connection that is connected to the output of the next weight layer with a ReLU activation function. When the output feature maps are of identical sizes, an equivalent number of filters is employed across layers. In the scenarios where the feature map size is reduced by half, the number of filters is augmented twofold.
The process of downsampling is performed by convolutional layers featuring a stride of 2, while batch normalization is applied immediately following each convolution. When input and output dimensions are the same, an identity shortcut mechanism is implemented. In the case when dimensionality is increased, a projection shortcut technique is employed by 1$\times$1 convolutions to match the dimension. The core innovation of ResNet is in its residual block which is shown in Figure \ref{fig:QTL_Architecture}(a), and can be represented as follows,
\begin{equation}
y=\sum F(x,{W_i})+x,
\label{eq:1}
\end{equation}

where $x$ is the input, $y$ is the output and $F(x,{W_i})$ represents the residual mapping to be learned with weight $W_i$. The last layer of ResNet-18 is a 512-feature map that extracts abstract and complex features from the input, followed by a fully connected layer. As the pre-trained ResNet-18 model is already trained on a dataset with a lot of diverse image categories, the pre-train network will act as a good feature extractor for our data (Ants/Bees, Road Sign Detection, CIFAR-10), even if the new data are from completely different classes. In the proposed QTL architecture, all the layers of ResNet-18 are frozen throughout the training except the last dense layer as shown in Figure \ref{fig:QTL_Architecture} (c). Thus, the weights of the pre-trained model are not updated during the training of our QTL model. The pre-trained model is used only as a feature extractor to extract the features from the input data. The last activation feature map of ResNet-18 outputs the bottleneck features which are then flattened before feeding to a fully connected layer. Here, we modified the output of the fully connected layer as six which is fed to the proposed quantum variational circuit. 
\vspace{-4mm}
\subsection{Quantum Classical Hybrid Convolutional Neural Network (QCHCNN)}
The proposed quantum circuit consists of three steps: data encoding, QVC (quantum variational circuit), and measurement. As the output of the last layer of the neural network is six, we use six qubits as the input of the quantum circuit for encoding classical data with six features. Each classical feature is mapped to a quantum state on a qubit. Here, each classical feature corresponds to a qubit. For data encoding, we followed the same encoding as followed by the Ref. \cite{qi2022classical, mari2020transfer}. To initiate the quantum states, the Hadamard gate is applied to the qubits to induce a superposition state. This superposition state signifies a combination of possible states that the qubits can assume. The superposition state is then subjected to the Rotational Y $(R_y)$ gate. This gate serves as the data embedding layer, acting as an intermediary between the quantum circuit and the preceding layer of the neural network. The $R_y$ gate manipulates the quantum states in a way that captures and encodes essential features from the data, thus facilitating the seamless integration of quantum and classical information within the hybrid model. If the classical input vector is $\theta = [\theta_1, \theta_2, \theta_3, \theta_4, \theta_5, \theta_6]$, quantum encoding helps to generate quantum embedding from this classical input. Hence, the quantum state can be represented as, 
\begin{align}
|\theta\rangle=|\theta_1\rangle \otimes |\theta_2\rangle \otimes |\theta_3\rangle \otimes |\theta_4\rangle\otimes |\theta_5\rangle\otimes |\theta_6\rangle \nonumber  \\
= \begin{bmatrix}
cos(\theta_1) \\ sin(\theta_1) \\
\end{bmatrix}  \otimes \begin{bmatrix}
cos(\theta_2) \\ sin(\theta_2) \\ 
\end{bmatrix} 
\begin{bmatrix}
cos(\theta_3) \\ sin(\theta_3) \\
\end{bmatrix}  \nonumber  \\ \otimes 
\begin{bmatrix}
cos(\theta_4) \\ sin(\theta_4) \\
\end{bmatrix}
\begin{bmatrix}
cos(\theta_5) \\ sin(\theta_5) \\
\end{bmatrix}  \otimes 
\begin{bmatrix}
cos(\theta_6) \\ sin(\theta_6) \\
\end{bmatrix} \nonumber  \\
= (\otimes_{i=2}^6 R_y(2x_i))|\theta_1\rangle^{\otimes^6}.
\label{eq:2}
\end{align}

The quantum encoding circuit in Figure \ref{fig:QTL_Architecture}(b) produces the following quantum state,

\begin{align}
(\otimes_{i=2}^6 R_y(2x_i))|\theta_1\rangle^{\otimes^6}
= \begin{bmatrix}
cos(\pi\theta_1) \\ sin(\pi\theta_1) \\
\end{bmatrix}  \otimes \begin{bmatrix}
cos(\pi\theta_2) \\ sin(\pi\theta_2) \\ 
\end{bmatrix} \nonumber  \\
\begin{bmatrix}
cos(\pi\theta_3) \\ sin(\pi\theta_3) \\
\end{bmatrix}  \otimes 
\begin{bmatrix}
cos(\pi\theta_4) \\ sin(\pi\theta_4) \\
\end{bmatrix}
\begin{bmatrix}
cos(\pi\theta_5) \\ sin(\pi\theta_5) \\
\end{bmatrix}  \otimes 
\begin{bmatrix}
cos(\pi\theta_6) \\ sin(\pi\theta_6) \\
\end{bmatrix}.
\label{eq:2}
\end{align}

A quantum variational Circuit (QVC) is designed to encode the optimization problem into a quantum circuit, where some of the gate parameters are trainable. These parameters are then optimized using classical optimization algorithms to find the optimal solution to the given problem. CNOT (Controlled-NOT) is a two-qubit gate for creating entanglement between qubits, where the state of one qubit becomes correlated with the state of another, even if they are physically separated. In QVC, there is a combination of CNOT gates and single-qubit adjustable rotation gates ($R_X$, $R_Y$, $R_Z$). The CNOT gates are used to entangle between qubits. Entanglement allows qubits to exhibit correlations that classical systems cannot achieve. The rotation angles $\alpha_i$, $\beta_i$, and $\gamma_i$ are the parameters associated with the adjustable rotation gates. These angles determine the amount of rotation around the X, Y, and Z axes, respectively. By adjusting these angles, the state of individual qubits can be manipulated, such as effectively preparing them in specific quantum states. These parameters are treated as trainable.

The quantum state outputs of the QVC circuit are projected by assessing the expectation values of six observables denoted as $z=[z_1,z_2,z_3,z_4z_5,z_6]$. These observables serve to characterize distinct properties of the quantum system. The goal is to ascertain the average values of these observables, which reflect the anticipated outcomes when measuring the same quantum state multiple times. Hence, the quantum state measurements into classical vectors can be represented as,
\begin{align}
M:|\theta \rangle \rightarrow z = \langle \theta|z|\theta \rangle,
\label{eq:2}
\end{align}
where M is the measurement operation. The QNN model uses a first-order optimisation technique to update the parameters such as stochastic gradient descent (SGD). The goal is to minimize the loss function over the dataset, thereby training the QNN to perform well on the given task. The first-order optimisation is expressed as,
\begin{align}
\theta(t+1) = \theta(t) - \eta \cdot \nabla L(\theta(t)),
\label{eq:5}
\end{align}
where $\theta(t)$ is the vector of model parameters at iteration $t$, $\theta(t+1)$ is the updated vector of model parameters at iteration $(t+1)$, $\eta$ is the learning rate, a hyperparameter that determines the step size for the update, $\nabla L(\theta(t))$ is the gradient of the loss function $L$ with respect to the model parameters. For a small value of $\epsilon$, the finite difference method is used to approximate the partial derivative as,
\begin{align}
\frac{\partial L(\theta)}{\partial \theta} \approx \frac{L(\theta + \varepsilon) - L(\theta)}{\varepsilon}.
\label{eq:2}
\end{align}
Finally, a widely used loss function in classical ML, the cross-entropy loss is used as a  classification loss to calculate the difference between the predicted probability distribution and the true probability distribution of the target classes. In the context of binary classification (two classes), the cross-entropy loss can be represented as,
\begin{align}
L_{classification} = -[ y \cdot \log(p) + (1 - y) \cdot \log(1 - p) ],
\label{eq:7}
\end{align}
where $y$ is the true label, $p$ is the predicted probability that the sample belongs to the positive class (i.e., class 1) and $1-p$ is the predicted probability that the sample belongs to the negative class (i.e., class 0). If $y=1$, the loss is only affected by $log(p)$, which encourages the predicted probability $p$ to be close to 1 which is a correct prediction for the positive class. If $y=0$, the loss is only affected by $-log(1-p)$, which encourages the predicted probability $p$ to be close to 0 which is the indication of a correct prediction for the negative class.

For multi-class classification, the cross-entropy loss can be defined as,
\begin{equation}
L_{classification}= - \sum_{i=1}^{n}p_ilog\bar{p_i}=-log\bar{p}_t,
\end{equation}
where $p_i$ is the probability distribution of the target, $i$ is the number of classes, $\bar{p}_i$ is the predicted probability distribution and  $t$ is the target class. 

\subsection{QCHCNN without Transfer Learning}
\label{without_TL}
In this section, to make a fair comparison of our proposed QTL approach, we develop an architecture of a hybrid classical-quantum method where there is no transfer learning involved. The Hybrid Classical-Quantum QNN architecture utilises a pre-trained ResNet-18 as a feature extractor to capture hierarchical features from the raw input data. Even though the ResNet-18 is pre-trained, the entire architecture is fine-tuned from scratch. This means that all the layers in the network, including from ResNet-18, are updated during the training process as illustrated in Figure \ref{fig:QTL_Architecture}(d). The classical part of the architecture includes initial layers that preprocess the input data. These layers encode the quantum data into a format suitable for subsequent quantum processing. The output of the pre-trained ResNet-18 is a set of high-dimensional features that capture meaningful information from the input data. Dealing with these high-dimensional complex features is computationally expensive and inefficient. Thus, the next step is to compress these features into a lower-dimensional space. The compressed high-dimensional features are connected to a Dense layer, which is a type of fully connected neural network layer. This layer learns relevant representations from the compressed features. The resulting features are then fed into the quantum encoding framework. After passing through the QNN, the quantum states are measured to obtain classical outputs. The QNN processes the data in quantum space and extracts unique and complex patterns that are challenging for classical methods to obtain. The measured outputs from the QNN are then connected to a classical fully connected layer using a non-trainable matrix. This last layer takes the quantum-processed information and combines it with classical processing to make the final predictions for the classification task. The non-trainable matrix ensures that the quantum information is integrated into the classical processing.
The training process of the Hybrid Classical-Quantum QNN involves optimizing the weights and parameters of both the classical pre-processing layers and the quantum circuit. This optimization is guided by a loss function that determines the discrepancy between the predicted labels and the ground truth labels. To update the parameters of the network, we employ gradient-based optimization techniques, adjusting the weights in a way that minimizes the loss function. Given the inherent complexity of quantum operations, training quantum circuits typically involves a degree of computational challenge. Quantum state optimization techniques, such as parameter shift and gradient-based methods, are employed to fine-tune the parameters of the quantum circuit. Unlike the QTL framework where transfer learning accelerates convergence and enhances performance, the hybrid classical-quantum QNN requires more iterations and computational resources for training, as the network learns both classical and quantum representations jointly.

\subsection{ Adversarial Attacks on Quantum Models}
\label{adversarial_attack}
\begin{figure*}
\begin{center}
\includegraphics[width=1.0\linewidth]{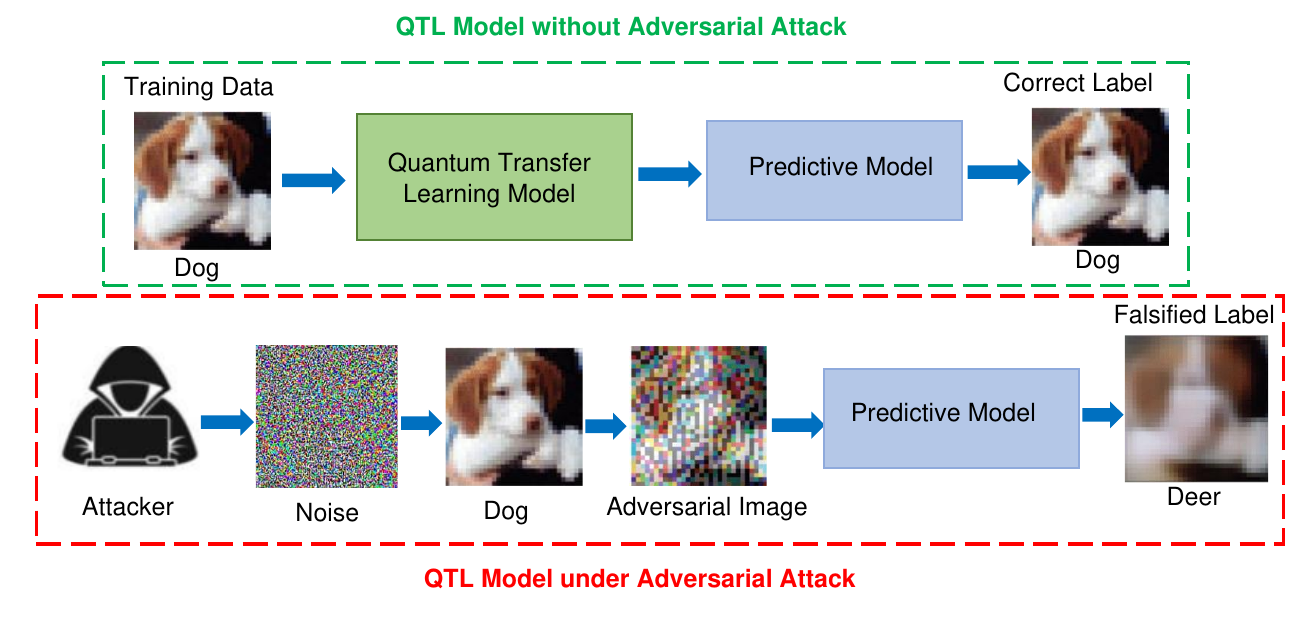}
\end{center}
   \caption{Overview of quantum adversarial attack. If we consider the input image of a dog, the QTL model correctly classifies it as a dog when the model is not under adversarial attack. However, a carefully crafted perturbation is applied to the input image in a way that exploits the vulnerabilities of the proposed QTL model. When the perturbed images are fed to the model for prediction, the QTL model incorrectly classified the image as a deer rather than a dog. This is an indication that the QTL model is vulnerable to small changes in input data, leading to significant implications for the reliability and security of the model.}
\label{fig:adversarial_attack}
\end{figure*}

Adversarial attacks are used to manipulate the predictions of the models, specifically the classifiers. Let us assume that $f:X \rightarrow Y$ is a classifier that maps input data $X$, which is often a subset of $R^d$, to discrete output labels $Y$. The objective of an adversarial attack is to craft an example $x^\wedge$ that is very close to a legitimate training sample $x$, denoted by $||x^\wedge - x ||$, while causing the classifier's prediction for 
$x^\wedge$ to differ from its prediction for $x$, i.e., $f(x^\wedge) \not = f(x)$. When an attack's goal is to make  $f(x^\wedge) \not = f(x)$, without any specific target label, it is referred to as a non-targeted attack. In contrast, in a targeted attack, the adversary aims to force the classifier's prediction for the adversarial example $x^\wedge$ to match a predefined target label $y$, i.e., $f(x^\wedge) = y$. In both cases, the difference between the adversarial example  $x^\wedge$ and the original example $x$ should be small, and this closeness is typically measured using the $L_p$ norm, where $p$ is usually chosen to be 0, 1, 2, or $\infty$. Different values of $p$ represent different ways of measuring the distance between two examples, such as the absolute differences in coordinates are measured for $L_1$ norm, the Euclidean distance is for $L_2$ norm, or the maximum absolute difference is for  $L_\infty$ norm. An adversarial attack method is illustrated in Figure \ref{fig:adversarial_attack}.

In our proposed architecture, we adopt the Fast Gradient Sign Method (FGSM) method for generating adversarial examples to attack the proposed QTL network. FGSM works by perturbing the input data in a way that maximizes the model's prediction error. The idea behind FGSM is to compute the gradient of the loss with respect to the input data and then perturb the data in the direction that increases the loss the most. For a classifier $f:X \rightarrow Y$, the goal of the FGSM attack is to generate an adversarial sample $x^\wedge$ that is similar to the original training example $x$. The FGSM method calculates the gradient of the loss function with respect to the input $x$ as,
\begin{align}
\Delta_x L(y,f(x;\theta)),
\label{eq:9}
\end{align}
where $y$ represents true label and $\theta$ represents models's parameters. This gradient indicates how the loss changes with small variations in the input $x$. The adversarial example $x^\wedge$ is generated by perturbing the original input $x$ in the direction that maximizes the loss,
\begin{align}
x^\wedge = x + \epsilon \cdot \text{sign}(\nabla_x L(y, f(x; \theta))),
\label{eq:9}
\end{align}
Here, $\epsilon$ is a small positive constant that determines the magnitude of the perturbation. The 
$sign$ function extracts the positive or negative direction of the gradient, indicating the direction in which to perturb the input sample.
\begin{figure*}
\begin{center}
\includegraphics[width=1.0\linewidth]{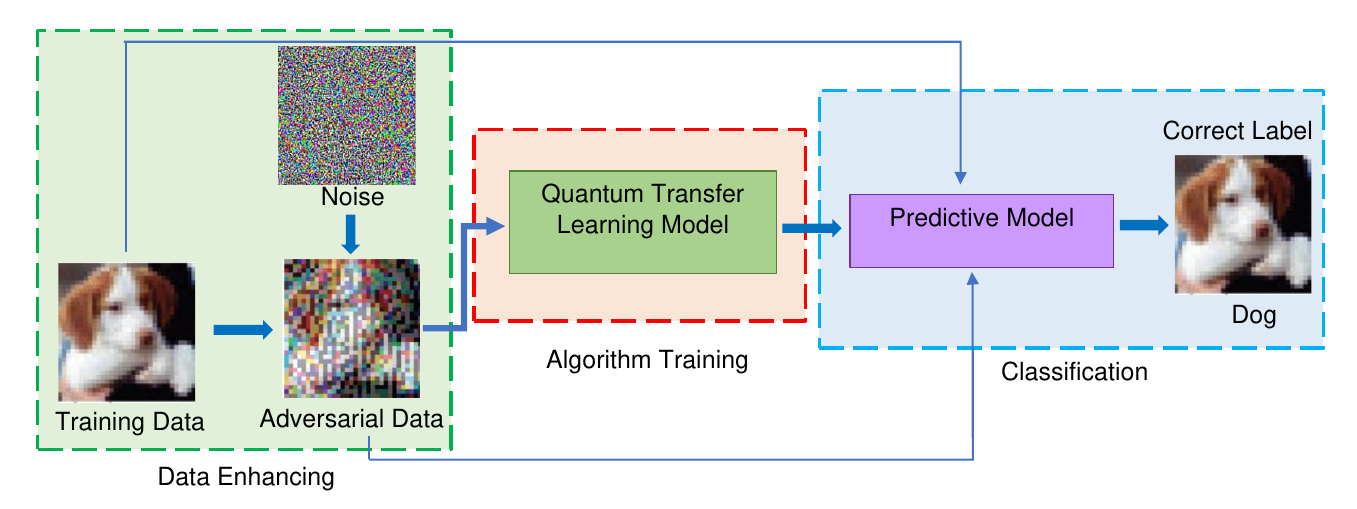}
\end{center}
   \caption{Illustration of quantum adversarial transfer learning architecture. For adversarial training, the training data now consists of both original input images and adversarial examples. As the QTL network is trained on this combined dataset, the network encounters both original and adversarial images. Thus, the network learns to resist the effects of adversarial perturbations and make accurate predictions even when presented with adversarial examples. As a result of adversarial training, the QTL network becomes more robust against adversarial attacks. When presented with a perturbed image that was crafted to deceive the network, it is now less likely to be misclassified. In the figure, where a dog image was misclassified as a deer, the adversarial training would help the network correctly classify it as a dog.}
\label{fig:adversarial_training}
\end{figure*}
\subsection{ Adversarial Training for Quantum Robustness}
\label{adversarial_training}
Adversarial training is employed in machine learning to enhance the robustness of neural networks against adversarial attacks. These attacks involve making small, carefully perturbed changes to input data to deceive a model's predictions. The overview of adversarial training is illustrated in Figure \ref{fig:adversarial_training} where the proposed hybrid QTL method is trained with adversarial images along with the original training images. Here, we consider a scenario where a QTL network is trained to classify images of animals. Adversarial attacks involve adding tiny, imperceptible changes to an image of a dog, for example, which would cause the QTL network to wrongly classify it as a deer. Adversarial training aims to improve the model's ability to resist such manipulations. In other words, the goal of adversarial training is to train models that can withstand such attacks and provide accurate predictions even in the presence of these perturbations. Adversarial training finds model parameters that minimise the maximum loss over all possible perturbations. Mathematically, this can be represented as,
\begin{align}
\min_{\theta} \max_{\delta \in \Delta} \mathcal{L}(f_{\theta}(x + \delta), y.
\label{eq:11}
\end{align}
Here, $\theta$ is the model parameter, where $x$ is the input data and $y$ is the corresponding correct output label, $f_{\theta}(x)$ is the predictions. The loss function $\mathcal{L}$ determines the difference between the predictions of the network $f_{\theta}(x)$ and the actual labels $y$. Minimizing this loss function during training helps the network to make accurate predictions. $\delta$ is the threat of the model that defines the types of potential attacks that the model might face. In this case, the threat model is defined as $\Delta = \{ \delta : \|\delta\|_{\infty} \leq \varepsilon \}$, where $\epsilon$ is a small positive value. Thus, the attack is limited to adding perturbations $\delta$ to the input data, and these perturbations are bounded by the maximum $\mathcal{L}_\infty$ norm of $\epsilon$. However, achieving the inner maximisation over the perturbation space $\Delta$ is challenging, thus, FGSM attack is used to approximate this maximisation. FGSM calculates the perturbation as,
\begin{align}
\delta^* = \varepsilon \cdot \text{sign}(\nabla_x \mathcal{L}(f(x), y)),
\label{eq:9}
\end{align}
where $\nabla_x$ represents the gradient with respect to the input data. The model parameters are updated using gradient descent, considering the approximated perturbation $\delta^*$ as calculated by the adversarial attack. After each gradient step, the perturbation $\delta$ pushes the input data outside the defined threat model. To ensure that the perturbation remains within the threat model, it is projected back into the set $\Delta$. This projection involves clipping $\delta$ to the interval $[-\epsilon, \epsilon]$. As the projection step ensures that perturbations stay within the allowed threat model, the adversarial training technique helps in training models that can resist adversarial attacks and maintain accurate predictions even in the presence of perturbations. 

\section{Network setup and training}
We use PennyLane \cite{bergholm2018pennylane} to train the QTL model and PyTorch \cite{paszke2019pytorch} for classical machine learning components. Adam optimizer is used for all the experiments with a batch size of 16 and a learning rate of 0.0004. The learning rate is constant for the first 25 epochs, and then linearly decays towards zero over the next 25 epochs. We initialise the quantum circuit with 6 qubits and 6 variational layers (depth of the quantum circuit). To enhance training convergence, a learning rate reduction strategy was employed, involving a reduction of 0.1 applied every 10 epochs. We initiate the training process with an initial spread of random quantum weights as 0.01.
\begin{figure*}
\begin{center}
\includegraphics[width=1.0\linewidth]{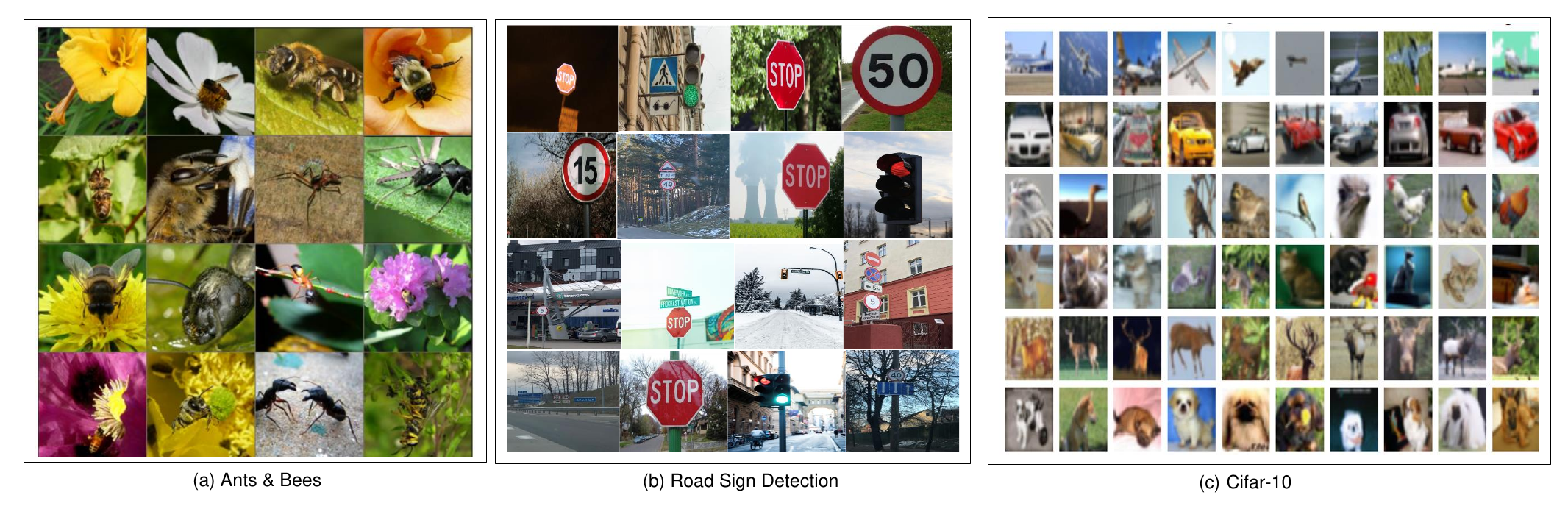}
\end{center}
   \caption{Overview of datasets used to evaluate the proposed QTL method. In the Figure, (a) represents Ants \& Bees dataset, (b) illustrates Road sign detection dataset and (c) represents CIFAR-10 \cite{krizhevsky2009learning} dataset. Each dataset encapsulates distinct challenges and contexts, collectively contributing to a comprehensive assessment of the QTL method's performance.}
\label{fig:Datasets}
\end{figure*}
\section{Experimental results and discussions}

\subsection{Datasets}
To evaluate the proposed QTL method and its robustness against adversarial attacks, three popular high-resolution image datasets (Ants \& Bees, CIFAR-10 and Road Sign Detection) are used. While existing QML/QAML studies mostly considered small-scale low-resolution image datasets such as MNIST (28$\times$28$\times$1) or CIFAR (32$\times$32$\times$3), our proposed method is evaluated on high-dimensional image datasets which are explained in the following section.

\textbf{Ants \& Bees} \cite{paszke2019pytorch} dataset is relatively small, containing a total of 120 images of ants and bees. The dataset also includes an additional 75 images for each class (ants and bees) for validation purposes. The image size is 768$\times$512 pixels. This dataset size is considered quite limited for training deep learning models from scratch, as complex models typically require a large amount of data to generalize well and learn meaningful patterns. However, our QTL model already pre-trained and learned features from a different, larger dataset. Thus, the advantage of the proposed model lies in the model's ability to leverage its pre-existing knowledge when adapting to the small-sized Ants \& Bees dataset.

\textbf{CIFAR-10 (Canadian Institute for Advanced Research-10)} \cite{krizhevsky2009learning} is a popular benchmark dataset in computer vision and machine learning. The dataset is designed for image classification tasks and is commonly used to evaluate and compare the performance of various machine learning models and algorithms. The CIFAR-10 dataset contains 60,000 color images in 10 different classes, with each class representing a different object or category. The training set contains 50,000 labeled images, with each class having an equal representation. The test set contains 10,000 labeled images, which are used to evaluate the performance of the models that are trained on the training set. Each image in the dataset is a 32$\times$32 color image, which means it has a height and width of 32 pixels and consists of three color channels (red, green, and blue). Despite its small image size and simplicity, this dataset presents challenges due to the relatively low resolution and the diversity of objects in the images. 

\textbf{Road Sign Detection} \cite{mogelmose2012vision} dataset comprises a total of 877 images, each with a resolution of 1024$\times$1024 pixels. This dataset is designed to identify and classify various types of road signs present in the images. There are four distinct classes, with each class representing a specific type of road sign: traffic lights, stop signs, various speed limit signs, and pedestrian crosswalks. In our experiments, this dataset is used to train and evaluate the proposed QTL method for road sign classification. Such models play a crucial role in autonomous driving systems, advanced driver assistance systems (ADAS), and other traffic-related applications. For our experiments, we split the dataset as 80/20 for training and testing the model.
\begin{table}[!t]
\fontsize{8}{8}\selectfont
\centering
\begin{tabular}{|p{3.4cm}|p{3cm}|p{1.5cm}|}
\hline
\textbf{Method} & \textbf{Dataset} & \textbf{Accuracy}\\
\hline\hline
Classical TL &Ants \& Bees &94.7\% \\
Classical without TL  &Ants \& Bees &62.75\% \\
Quantum TL  &Ants \& Bees &96.1\%\\
Quantum without TL  &Ants \& Bees &65.36\% \\
\hline
Classical TL &CIFAR-10 &92.1\% \\
Classical without TL  &CIFAR-10 &72.9\% \\
Quantum TL  &CIFAR-10 &95.8\%\\
Quantum without TL  &CIFAR-10 &51.7\% \\
\hline
Classical TL &Road Sign Detection &92.47\% \\
Classical without TL  &Road Sign Detection &60.22\% \\
Quantum TL  &Road Sign Detection &94.62\%\\
Quantum without TL  &Road Sign Detection &48.39\% \\
\hline
\end{tabular}
\caption{Simulation results of the proposed QTL approach compared to the classicalTL method on Ants \& Bees \cite{paszke2019pytorch}, CIFAR-10 \cite{krizhevsky2009learning}, and Road Sign Detection \cite{mogelmose2012vision} datasets. The performance are also reported for both classical and quantum when no TL is involved.}
\label{tab:TL_results}
\end{table}
\subsection{Comparison of QTL with classical TL approach}

\begin{table*}[!htbp]
\begin{center}
\fontsize{9}{9}\selectfont
\centering
\begin{tabular}{|p{3.5cm}|p{1.8cm}|p{1.8cm}|p{1.8cm}|p{2.2cm}||p{2.2cm}|}
\hline
\textbf{Method} & \textbf{Attack Strength} & \textbf{Clean \newline Accuracy} & \textbf{Accuracy under  \newline Attack} & \textbf{Clean \newline Accuracy with \newline adversarial data} & \textbf{Adversarial Training  \newline Accuracy}\\
\hline\hline
Classical TL & 0.1 &94.70\% &50.64\% &92.38\% &64.10\% \\
Classical TL & 0.2 &94.70\% &47.70\% &89.17\% &61.10\% \\
Classical TL & 0.3 &94.70\% &45.79\% &86.93\% &59.10\% \\
Classical without TL & 0.1 &62.75\% &44.87\% &59.48\% &47.64\% \\
Classical without TL & 0.2 &62.75\% &44.50\% &57.54\% &47.01\% \\
Classical without TL & 0.3 &62.75\% &44.23\% &56.44\% &46.79\% \\
\hline
Quantum TL & 0.1 &96.10\% &53.85\% &94.76\% &71.87\% \\
Quantum TL & 0.2 &96.10\% &50.90\% &94.53\% &69.30\% \\
Quantum TL & 0.3 &96.10\% &47.45\% &94.12\% &67.82\% \\
Quantum without TL & 0.1 &65.36\% &45.68\% &49.23\% &48.12\% \\
Quantum without TL & 0.2 &65.36\% &45.12\% &48.74\% &48.0\% \\
Quantum without TL & 0.3 &65.36\% &45.12\% &48.06\% &47.98\% \\
\hline
\end{tabular}
\end{center}
\caption{Simulation results of the proposed QTL approach on Ants \& Bees dataset under adversarial attack with a different attack strength are compared to the accuracy of classical TL under attack. The experiments are also performed on AT to evaluate the robustness of the proposed method.}
\label{tab:Ants_bees_AT}
\end{table*}

The results of our proposed QTL on all three datasets are shown in Table \ref{tab:TL_results}. We also compare our results with their classical counterparts for fair evaluation. For Ants \& Bees dataset, the proposed method achieves 96.1\% accuracy while the accuracy is 94.7\% for classical TL. For CIFAR-10 and Road sign detection datasets, we achieved 95.8\% and 94.6\% accuracy while classical TL achieves 92.1\% and 92.4\%, respectively. It is noted that the proposed QTL method outperforms the classical TL method by 1.4\%, 3.7\%, and 2.2\% classification accuracy for Ants \& Bees, CIFAR-10 and Road sign detection datasets, respectively. We also perform experiments on both classical CNN and QNN without TL to further justify the performance of the proposed QTL
method, i.e., when there is no TL involved and the network is trained from scratch. For these experiments, we followed the architecture illustrated in Figure \ref{fig:QTL_Architecture}   and Section
\ref{without_TL}. The results of CNN and QNN without TL are reported in Table \ref{tab:TL_results}. From Table \ref{tab:TL_results}, it is noted that the classification accuracy of quantum without TL is 65.36\% for the Ants \& Bees dataset, which is a 30.74\% decrease in accuracy than the QTL approach. Similar deterioration can be observed for CIFAR-10 and Road sign detection datasets, i.e., 51.7\% and 48.39\% classification accuracy of QNN without TL, indicating a decrease of 44.1\% and 46.23\% compared to quantum with TL method. The comparison indicates the effectiveness of TL in QNNs.

\subsection{Adversarial Attack Evaluation and Robustness through Adversarial Training}
\label{AT_results}
The robustness of the proposed QTL approach is evaluated by subjecting the model to adversarial attack. The adversarial attack is generated as explained in Section \ref{adversarial_attack}. The results for Ants \& Bees dataset are reported in Table 
\ref{tab:Ants_bees_AT}. Classical TL, classical without TL, quantum TL, and quantum without TL models is evaluated with various attack strengths ($\epsilon$) such as 0.1, 0.2, and 0.3. The clean accuracy of the model (before the adversarial attack), the accuracy when the model is under attack, and the clean accuracy with adversarial data are reported in Table \ref{tab:Ants_bees_AT}. We observed that the accuracy of the proposed QTL drops by 42.25\% when the model is attacked with an attack strength of 0.1 while the clean accuracy is 96.1\%. When the attack strength is increased to 0.2, a further 2.95\% accuracy drop is reported. The deterioration in classification accuracy is a clear indication that the quantum TL model is as vulnerable to adversarial attack as the classical TL method. As we proposed to enhance the robustness of the QTL model against adversarial attacks by introducing AT (see Section \ref{adversarial_training}) within the proposed QTL approach, the results of the adversarially trained model are reported in Table \ref{tab:Ants_bees_AT}. When the QTL model has trained adversarially with attack strengths of 0.1, 0.2, and 0.3, we achieve 71.87\%, 69.30\%, and 67.82\% accuracy respectively, indicating a significant improvement in accuracy. The model achieved performance gains such as 18.02\%, 18.4\%, and 20.35\% increase in accuracy compared to the performance when the model was under attack. The results of QNN without TL are also reported for further clarification. We also noticed that, when no TL is involved, the performance of the classical network is also significantly dropped, indicating the effect of TL to improve the performance of the model. It is also noted from a comprehensive review article Ref. \cite{ijcai2021p591} that the state-of-the-art classical CNN model was also heavily affected due to adversarial attack.

\begin{table}[!htbp]
\begin{center}
\fontsize{8}{8}\selectfont
\centering
\begin{tabular}{|p{1.8cm}|p{1.3cm}|p{1.4cm}|p{1.4cm}|p{1.8cm}|}
\hline
\textbf{Method} & \textbf{Attack Strength} & \textbf{Clean \newline Accuracy} & \textbf{Accuracy under Attack} 
& \textbf{Adversarial \newline Training \newline Accuracy}\\
\hline\hline
Classical TL & 0.1 &93.7\% &85.4\%  &93.7\% \\
Classical TL & 0.3 &93.7\% &83.3\%  &92.4\% \\
Classical TL & 1.0 &93.7\% &58.7\% &89.5\% \\
\hline
Quantum TL & 0.1 &95.8\% &89.6\% &95.8\% \\
Quantum TL & 0.3 &95.8\% &87.5\%  &93.7\% \\
Quantum TL & 1.0 &95.8\% &60.4\%  &92.1\% \\
\hline
\end{tabular}
\end{center}
\caption{Simulation results of the proposed QTL approach on Road Sign detection dataset under adversarial attack with a different attack strength is compared to the accuracy of classicalTL under attack. The experiments are also performed on AT to evaluate the robustness of the proposed method.}
\label{tab:road_data}
\end{table}

\begin{figure}
\includegraphics[width=1.0\linewidth]{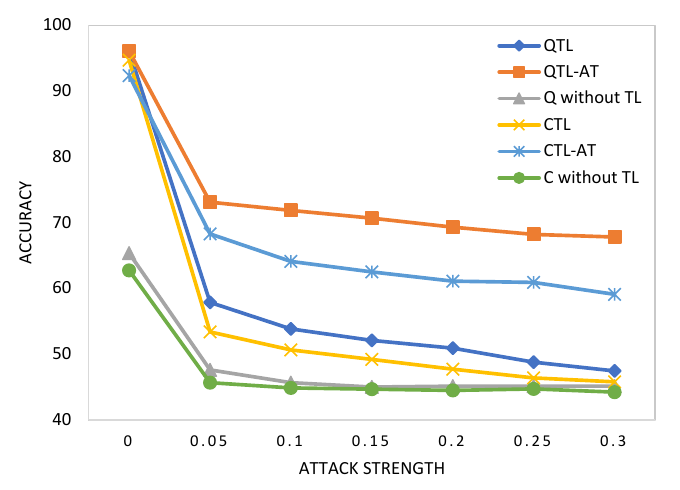}
   \caption{The plot of accuracy versus attack strength on Ants \& Bees for QTL (quantum transfer learning), QTL-AT (quantum transfer learning with adversarial training), Q without TL (quantum without transfer learning), CTL (classical transfer learning), CTL-AT (classical transfer learning with adversarial training), C without TL (classical without transfer learning).}
\label{fig:graph}
\end{figure}

\begin{table}[!htbp]
\begin{center}
\fontsize{8}{8}\selectfont
\centering
\begin{tabular}{|p{1.8cm}|p{1.3cm}|p{1.4cm}|p{1.4cm}|p{1.8cm}|}
\hline
\textbf{Method} & \textbf{Attack Strength} & \textbf{Clean \newline Accuracy} & \textbf{Accuracy under Attack} & \textbf{Adversarial \newline Training \newline Accuracy}\\
\hline\hline
Classical TL & 0.1 &92.1\% &50.1\% &85.8\% \\
Classical TL & 0.3 &92.1\% &50.2\% &84.9\% \\
Classical TL & 1.0 &92.1\% &49.2\% &80.8\% \\
\hline
Quantum TL & 0.1 &95.8\% &50.9\% &86.6\% \\
Quantum TL & 0.3 &95.8\% &50.0\%  &86.2\% \\
Quantum TL & 1.0 &95.8\% &50.0\% &85.0\% \\
\hline
\end{tabular}
\end{center}
\caption{Simulation results of the proposed QTL approach on CIFAR-10 dataset under adversarial attack with a different attack strength is compared to the accuracy of classical TL under attack. The experiments are also performed on AT to evaluate the robustness of the proposed method.}
\label{tab:cifar-10}
\end{table}
For Road Sign Detection and CIFAR-10, the adversarial attack and AT results are reported in Table \ref{tab:road_data} and \ref{tab:cifar-10}, respectively. For the Road Sign Detection dataset, when the QTL model is attacked with a strength of 0.1, a decrease in performance is noticed from 95.8\% to 89.6\%. However, we noticed that even when the attack strength is 0.3, the accuracy is only dropping 2.1\%. Hence, we increase the attack strength to 1.0 and notice a significant reduction of model performance to 60.4\%. For CIFAR-10, we also evaluate the performance of the proposed model with the attack strength of 0.1, 0.3 and 1.0 which affects the model accuracy to drop down from 50.9\% to 50.0\%. It is observed that even if the attack strength is 1.0, the accuracy is not going below 50.0\% for CIFAR-10. AT is also performed for both Road Sign Detection and CIFAR-10 datasets to make the model robust against adversarial attacks. In both cases, significant performance improvements are achieved, while the classification accuracy with AT is very close to the clean results for Road Sign Detection dataset. Additionally, Figure \ref{fig:graph} shows the plot of accuracy versus attack strength on Ants \& Bees dataset for QTL (quantum transfer learning), QTL-AT (quantum transfer learning with adversarial training), quantum without transfer learning, CTL (classical transfer learning), CTL-AT (classical transfer learning with adversarial training), classical without transfer learning. It is clear from the figure that QTL-AT outperforms all other methods, indicating the superior performance of the proposed method with AT.

\subsection{Discussion}
The results presented in our work demonstrate the effectiveness of the proposed QTL approach in leveraging the benefits of quantum computing and classical machine learning for image classification tasks. The QTL approach demonstrated better performance on multiple datasets, including Ants \& Bees, CIFAR-10, and Road Sign Detection, outperforming classical TL and QNNs without TL. One of the notable strengths of the proposed QTL approach is its ability to bridge the gap between the limited quantum resources, such as the number of qubits, and the high-dimensional nature of image data. The integration of a pre-trained classical CNN, ResNet-18, as a feature extractor in the quantum model contributes to this bridging by enabling the extraction of relevant and meaningful features from the raw image data. This knowledge transfer from a large-scale classical CNN to a small-scale Quantum Variational Circuit-based Quantum Neural Network leads the proposed QTL to achieve gain in performance. The achieved classification accuracies demonstrate the advantages of the QTL approach. The comparison of QTL with classical TL and QNN without TL clearly shows the superiority of QTL in terms of classification performance. For instance, on the Ants \& Bees dataset, the proposed QTL approach achieved an accuracy of 96.1\%, outperforming classical TL by 1.4\% and QNN without TL by a significant margin of 30.74\%. Similar trends are observed on CIFAR-10 and Road Sign Detection datasets, further confirming the robustness and effectiveness of the model. Furthermore, the evaluation of the QTL approach under adversarial attacks provides insights into its resilience and vulnerability. The proposed adversarial training enhances the model's ability to withstand adversarial perturbations, as evidenced by the improved accuracy when compared to non-adversarial trained models as discussed in Section \ref{AT_results}. The achieved improvements in classification accuracy with AT demonstrate the successful integration of robustness-enhancing techniques in the QTL framework.
\vspace{1mm}
\section{Conclusion}
In this paper, we present an end-to-end QTL approach that exploits the strengths of quantum computing and classical machine learning in a unified framework. Through comprehensive simulations on diverse datasets, including Ants \& Bees, CIFAR-10, and Road Sign Detection, we have demonstrated performance gain of our QTL approach compared to traditional classical TL methods. Our findings revealed that the QTL approach outperformed classical TL and QNNs without TL for all three datasets, showcasing its potential to significantly improve classification accuracy, even in scenarios with limited training data. Moreover, we address the the vulnerability of quantum models to adversarial attacks which is unexplored in QTL. By subjecting the QTL model to adversarial perturbations, we shed light on the potential security threats. To enhance the robustness of the proposed QTL model, we incorporated AT within the QTL framework and achieved improved performance, showcasing the effectiveness and reliability of the model. Our simulation results and extensive comparison of QTL with its classical counterparts open up avenues for future research in quantum-enhanced machine learning. 

\section*{Acknowledgments}
The authors acknowledge the use of CSIRO HPC (High-Performance Computing) and NCI's Gadi supercomputer for conducting the experiments. A.K. also acknowledges CSIRO's Quantum Technologies Future Science Platform for providing the opportunity to work on quantum machine learning.
\bibliographystyle{unsrt} 
\bibliography{apssamp}

\end{document}